\documentclass[fleqn,twoside]{article}
\usepackage{espcrc2}

\newcommand{\bm}[1]{\mbox{\boldmath $#1$}}

\newcommand{\bkt}{\bm k_{_T}}

\newcommand{\ba}{\begin{eqnarray}}
\newcommand{\ea}{\end{eqnarray}}
\newcommand{\beq}{\begin{equation}}
\newcommand{\eeq}{\end{equation}}

\newcommand{\nn}{\nonumber}

\newcommand{\text}[1]{\mbox{$\rm{#1}$}}

\newcommand{\AmS}{{\protect\the\textfont2
  A\kern-.1667em\lower.5ex\hbox{M}\kern-.125emS}}

\title{\mbox{}\\[-1 cm]
Theoretical Aspects of Transversity Observables}

\author{Dani\"{e}l Boer\address{Dept.\ of Physics and Astronomy, 
Vrije Universiteit\\ 
De Boelelaan 1081, 1081 HV Amsterdam, The Netherlands}\thanks{Talk 
presented at the
``International Workshop on the Spin Structure of the Proton and
Polarized Collider Physics'', ECT$^*$, Trento, Italy, 
July 23-28, 2001.}}
       
\begin{document}

\begin{abstract}
Theoretical aspects of transversity observables are reviewed. The main 
focus is on two leading twist transversity single spin asymmetries, one 
arising from the Collins effect and one from the interference fragmentation 
functions. Electron-positron annihilation experiments which are required to 
obtain these fragmentation functions are discussed, as well as the 
issues of factorization, evolution and Sudakov factors for the relevant 
observables. These theoretical considerations pinpoint the most realistic
scenarios towards measurements of transversity.
\end{abstract}

\maketitle

\section{Transversity}

\noindent
The transversity distribution function \cite{Ralst-S-79} 
$h_1$ (or $\delta q$)
is a measure of how much of the transverse spin of a polarized proton is 
transferred to its quarks, i.e.\
the density of transversely polarized quarks
inside a transversely polarized proton. 

The transversity distribution function 
is a chiral-odd or helicity flip amplitude. Observables involving 
transversity should therefore be (helicity flip)$^2$. This is the reason $h_1$
cannot be measured in inclusive Deep Inelastic Scattering (DIS) 
($e p \to e' X$); it enters the cross section suppressed by a 
factor of $m_a/Q$, where $m_a$ is the mass of a quark of flavor $a$ and 
$Q$ is the invariant mass of the virtual photon that probes the proton.
A further complication is that in charged current exchange processes 
chiral-odd functions like $h_1$ cannot be accessed. This is a drawback
regarding a future flavor separation for such functions.

To measure transversity there are essentially 
two options left: single or double transverse spin asymmetries in
(semi-inclusive, neutral current) $ep$ or $pp$ processes. 
Few such experiments have been performed to date and this is the reason
that no experimental data on $h_1$ is available thus far. 
Hints of nonzero $h_1$ 
come only from the HERMES \cite{HERMES} and E704 \cite{BravarDNN} 
experiments (the latter is a measurement of $D_{NN}$, see below). 
But a number of future experiments (e.g.\ HERMES,COMPASS,RHIC) are expected to
provide detailed information on transversity functions. 

\section{Transversity asymmetries}

\noindent
The Drell-Yan process of two colliding transversely polarized hadrons
producing a lepton pair was originally thought to be the best way of accessing
the transversity distribution, for instance at RHIC.
This double transverse spin asymmetry $A_{TT}^{DY}$ is proportional to
$h_1^a(x_1) \; \overline h{}_1^a(x_2)$.
The problem is that $h_1$ for antiquarks ($h_1^{\bar a} = \overline h{}_1^a$) 
inside a proton is presumably much 
smaller than for quarks and the asymmetry is not expected to be large. 
In fact, by using Soffer's inequality ($|h_1(x)| \leq {\scriptstyle
\frac{1}{2}} \left[ f_1(x) + g_1(x) \right]$), $A_{TT}^{DY}$ has been shown
\cite{Martin} to be 
small at RHIC, just beyond the experimental reach (but a future upgrade would
be very promising). 

Also the double transverse spin asymmetry in jet production
(directly proportional to $(h_1)^2$ at high $p_{_T}$) poses experimental
problems, because of the tiny cross sections one is dealing with. In Ref.\
\cite{WernerFut} it is shown that for RHIC the statistical error is not a 
problem, but due to the small values of the asymmetry, the systematic errors 
need to be under extremely good control. 

Another possible way to access the transversity distribution function via a 
double transverse spin asymmetry, involves the 
transversity {\em fragmentation\/} function $H_1$. It
measures the probability of
$q(s_{_T}) \to h(S_{_T}) + X$, where $h$ is a spin-1/2 hadron, for instance a 
$\Lambda$. The double transverse spin asymmetry $D_{NN}$ --the tranverse
polarization transfer-- involving both $h_1$ and 
$H_1$ occurs in the processes $e \; p^\uparrow \to e' \;  
\Lambda^\uparrow \; X$ and $p \; p^\uparrow \to 
\Lambda^\uparrow \; X$. The latter observable has been measured by E704
\cite{BravarDNN} and found to
be sizeable, but a solid conclusion about $h_1$ cannot be drawn due to the
rather low $p_{_T}$ range, which casts doubt on the use of a factorized
expression for the cross section. Furthermore, $H_1$ is also 
unknown and although it could be extracted from the double
transverse spin asymmetry in $e^+ \: e^- \to \Lambda^\uparrow \; 
\overline{\Lambda}{}^\uparrow \: X$, this will also pose quite a challenge.

In short, double transverse 
spin asymmetries do not seem promising to extract the 
transversity distribution in the near future. This leaves the  
option of single spin asymmetries (SSA), which all exploit fragmentation
functions of some sort. There are three options: 
1) measuring the transverse momentum of the final state 
hadron compared to the jet axis; 2) producing final state hadrons with 
higher spin, e.g.\ $\rho$ (related to interference fragmentation functions to
be discussed below);  
3) higher twist asymmetries which are suppressed by inverse
powers of the hard scale $Q$. The third option does not seem very promising.
 
\section{Collins effect asymmetries}

\noindent
The Collins effect refers to a nonzero correlation between 
the transverse spin $\bm s_{_T}$ of a fragmenting quark and the distribution 
of produced hadrons. More specifically, a transversely polarized quark can 
in principle 
fragment into particles (with nonzero transverse momentum $\bkt$) having a 
$\bkt \times \bm s_{_T}$ angular distribution around the jet axis or, 
equivalently, the quark momentum, see Fig.~1 of Ref.\
\cite{DB-DIS01}. The Collins
effect will be denoted by a fragmentation function $H_1^\perp(z,\bkt)$ and
if nonzero, it can lead to SSA in 
$e \, p^\uparrow \to e' \pi \, X$ and $p \, p^\uparrow \to \pi \, X$. 
There are some experimental indications that the Collins effect is indeed 
nonzero, e.g.\ SSA measured by HERMES \cite{HERMES,DB-99} 
and SMC \cite{Bravar} at relatively low energies. Also, SSA in 
$p \, p^\uparrow \to \pi \, X$ as measured by E704 and at the AGS (BNL) can
be (at least partially) 
explained as arising from a nonzero Collins function \cite{Boglione}.

\subsection{Collins effect in semi-inclusive DIS}

\noindent
Collins \cite{Collins-93} considered semi-inclusive DIS (SIDIS) 
$e \; p^\uparrow
\to e' \; \pi \; X$, where the spin of the proton is orthogonal to the 
direction of the virtual photon $\gamma^*$ and one observes the pion
transverse momentum $\bm P^{\pi}_{\perp}$, which has 
an angle $\phi^e_\pi$ compared to the lepton scattering plane. 
Collins has shown that the cross section for this process has an asymmetry 
that is proportional to the transversity function: $A_{T} \propto  
\sin(\phi^e_\pi + \phi^e_{_S}) \; |\bm S_{T}^{}| \; h_1 \; H_1^\perp$.
To discuss this SSA further, we will first project it out from the cross 
section (cf.\ Ref.\ \cite{Boer-Mulders-98}). Consider the cross sections 
integrated, but weighted with a function $W = W(|\bm
P^{\pi}_{\perp}|,\phi_{\pi}^e)$:
\beq
\langle W \rangle
\equiv \int d^2\bm P^{\pi}_{\perp}
\ W\,\frac{d\sigma^{[e \, p^\uparrow \rightarrow e' \,
\pi \, X]}} {dx\,dy\,dz\,d\phi^e\,d^2\bm P^{\pi}_{\perp}},
\eeq
where we restrict to the case of $|\bm P^{\pi}_{\perp}|^2 \ll Q^2$.  
We will focus on 
\ba
{\cal O} &\equiv &\frac{\big\langle 
\sin( \phi_{_C} ) \, |\bm P^{\pi}_{\perp}|  
\big\rangle}{{\scriptstyle 
\left[4\pi\,\alpha^2\,s/Q^4\right]}M_\pi}\nn \\
& = &
\vert \bm S_{T}\vert\,{\scriptstyle (1-y)} 
\sum_{a,\bar a} e_a^2
\,x\,h_{1}^{a}(x) z H_1^{\perp (1) a}(z),
\label{observableO}
\ea
where $\phi_{_C} = \phi^e_\pi+\phi^e_{_S}$ and  
\beq
H_1^{\perp (1)}(z) \equiv \int 
d^2 \bkt \frac{\bkt^2}{2 z^2 M_\pi^2} H_1^{\perp}(z,\bkt^2).
\eeq
At present all phenomenological studies of the Collins effect are performed 
using such tree level expressions. 
But the leading order (LO) evolution equations 
are known for $h_1$ (NLO even) and $H_1^{\perp (1)}$ 
(at least in the large $N_c$ limit \cite{Henneman}), showing that  
both functions evolve autonomously (and vanish asymptotically).  
This provides one with the LO $Q^2$ behavior of the observable ${\cal O}$, 
which arises solely from the LO 
evolution of $h_1$ and $H_1^{\perp (1)}$. This is however 
a nontrivial result, 
since this semi-inclusive process is not a case where collinear factorization
applies. In the differential cross section $d\sigma/d^2\bm P^{\pi}_{\perp}$ 
itself, beyond tree level soft gluon corrections do not cancel, such that 
Sudakov 
factors need to be taken into account and a more complicated factorization 
theorem applies \cite{CS-81,DB-01}. In fact, the observable ${\cal O}$ 
(Eq.\ (\ref{observableO})) is the only $|\bm P^{\pi}_{\perp}|$-moment 
of the Collins asymmetry in the cross section, that is not sensitive 
to the Sudakov factors. This observable would therefore be better 
suited for a LO analysis than the full $|\bm P^{\pi}_{\perp}|$-dependent 
asymmetry. 

Now we will look at the latter 
in the explicit example of the Collins effect asymmetry in SIDIS; more
specifically, $e \, p \to e'
\, \gamma^*(\bm{q}_{T}^{}) \, p \to e' \, \pi \, X$ ($\bm{q}_{_T}^{} = -z \bm
P^{\pi}_{\perp}$ and $\bm{q}_{_T}^2 \equiv Q_T^2 \ll Q^2$)
\beq
\frac{d\sigma(e\, p \to e' \pi X)}{dx dz dy d\phi_e d^{\,2}{\bm
q_{T}^{}}}
\propto 
\left\{ 1 + |\bm S_{T}^{}| \sin(\phi_{_C}) 
A(\bm{q}_{_T}^{}) \right\}.
\eeq
To get an idea about the
effect of Sudakov factors, we have assumed Gaussian transverse 
momentum dependence for $H_1^\perp (z, \bkt) $, such that the asymmetry's 
analyzing power $ A(\bm{q}_{_T}^{})$ is given by 
\beq
A(\bm{q}_{_T}^{}) = \frac{
\sum_{a}\;e_a^2 \; b(y)\; h_1^a(x) H_1^{\perp (1) a}(z)
}{\sum_{b}\; e_b^2 \; a(y)\;f_1^b(x) D_1^b(z)} {\cal
A}(Q_T),
\eeq
where $a(y)= (1-y+\frac{1}{2} y^2), b(y)=(1-y)$.
In Ref.\ \cite{DB-01} we studied ${\cal
A}(Q_T)$ for a generic nonperturbative Sudakov 
factor, because of lack of experimental input for this quantity.
It was found that ${\cal A}(Q_T)$ at $Q=M_Z$ becomes considerably smaller and 
broader than the tree level expectation. We also observed that 
$\max [{\cal A}(Q_T)] \sim Q^{-0.5} - Q^{-0.6}$.  
Thus, tree level estimates tend to overestimate
transverse momentum dependent azimuthal spin asymmetries and 
Sudakov factors cannot be ignored at present-day collider energies.

\subsection{Collins effect in $e^+ \, e^- \to \pi^+ \, \pi^- \, X$} 

\noindent
In order to obtain the Collins function itself,
one can measure a $\cos(2\phi)$
asymmetry in $e^+ \, e^- \to \pi^+ \, \pi^- \, X$, that has a contribution 
proportional to the Collins function squared \cite{BJM-97} (at equal
momentum fractions). 
A first indication of such a nonzero (but small) asymmetry comes from a 
preliminary analysis \cite{EST} of the 91-95 LEP1 data ({\small
DELPHI}). A similar study using the off-resonance data from the
B-factory BELLE at KEK, is planned \cite{Perdekamp}.

Also for this Collins effect observable, the tree level 
asymmetry expression is not sufficient if results from different experiments
are to be compared. Beyond tree level Sudakov factors 
need to be included. If the differential cross section is written as 
\beq
\frac{d\sigma (e^+e^-\to \pi^+ \pi^- X)}{d\Omega dz_1 dz_2 d^2{\bm
q_{_T}^{}}} 
\propto \left\{ 1 + \cos(2\phi) A(\bm{q}_{_T}^{}) \right\},
\eeq
with $\bm{q}_{_T}^2 \ll Q^2$, then assuming again Gaussian
transverse momentum dependence for the Collins function, we find
\beq
A(\bm{q}_{_T}^{}) \propto  \frac{H_1^{\perp (1)}(z_1) \overline H{}_1^{\perp
(1)}(z_2)}{D_1(z_1) \overline D{}_1(z_2) } {\cal A}(Q_T). 
\eeq
Again, a generic example \cite{DB-01} shows that Sudakov factors now   
produce an order of magnitude suppression at $Q=M_Z$ compared to a typical
tree level result (in addition, now $\max [{\cal A}(Q_T)] 
\sim Q^{-0.9} - Q^{-1.0}$). Therefore, this Collins 
effect observable is best studied with two jet events at $\sqrt{s} \ll M_Z$
(a requirement satisfied by BELLE, which operates on and just below 
the $\Upsilon(4S)$). 

Nevertheless, the extraction of the Collins function from this asymmetry is 
not straightforward, since there is asymmetric background from hard gluon 
radiation (when $Q_T \sim Q$) and from weak decays. The former enters the
$Q_T$ dependent asymmetry proportional to $\alpha_s Q_T^2/Q^2$, which 
at lower values of $Q^2$ need not be small. This
contribution could be neglected at LEP energies 
\cite{BJM-97}. Luckily it is calculable and so is 
the background from weak decays,
e.g.\ $e^+ e^- \to \tau^+ \tau^- \to \pi^+ \pi^- X$. 

As in the case of the Collins asymmetry in SIDIS, there is one particular
$Q_T$ moment of the asymmetry that is not sensitive to Sudakov factors, 
namely the first $Q_T^2$ moment: $\int dQ_T^2 Q_T^2 d\sigma/dQ_T$.
Unfortunately, it is mostly 
sensitive to the high $Q_T^2 (\sim Q^2)$ hard gluon radiation. 
This contribution could in principle be cut off
by imposing a maximum $Q_T$ cut, but this introduces a further source of 
uncertainty.  

\section{Interference fragmentation functions}

\noindent
Jaffe, Jin and Tang \cite{JJT} pointed out the possibility that the Collins 
effect averages to zero in the sum over final states $X$. 
Instead, they proposed to measure two pions in the final state
$\vert (\pi^+ \, \pi^-)_{{\text{out}}} X \rangle $ ($\pi^+, \pi^-$ 
belong to the same jet), which presumably depends on the strong phase 
shifts of the $(\pi^+ \, \pi^-)$ system. The interference between different 
partial waves could give rise to a nonzero chiral-odd fragmentation function 
called the interference fragmentation function (IFF). 
The IFF would lead to single spin asymmetries in 
$ e \, p^\uparrow \rightarrow e' \, (\pi^+ \, \pi^-) \, X$ and 
$p\, p^\uparrow \to (\pi^+ \, \pi^-) \, X$, both proportional to the
transversity function. 

\subsection{IFF in semi-inclusive DIS}

The SSA expression for 
$e \, p^\uparrow \rightarrow e' \, (\pi^+ \, \pi^-) \, X$ in terms of the IFF
$\delta\hat q_I^{}(z)$ is \cite{JJT} 
\beq
\big\langle \cos( \phi_{S_T}^e + \phi_{R_T}^e) 
\big\rangle \propto  F    
\vert \bm S_{T} \vert \vert 
\bm R_{T}\vert h_{1}(x) \delta\hat q_I^{}(z), 
\label{JJTasym}
\eeq
where $z=z^+ + z^-$; $\bm R_{T} 
= (z^+ \bm k^- - z^- \bm k^+)/z$; $F = F(m^2) = \sin \delta_0 \sin \delta_1
\sin(\delta_0-\delta_1)$, where $\delta_0, \delta_1$ are the $\ell = 0, 1$ 
phase shifts and $m^2$ is the $\pi^+ \pi^-$ invariant 
mass.
Note the implicit assumption of factorization of $z$ and $m^2$ 
dependence, which leads to the prediction that on the $\rho$
resonance the asymmetry is zero (according to the
experimentally determined phase shifts). More general $z, m^2$ 
dependences have been considered \cite{Bianconi} and this should be 
tested. 

Like-sign  
$(\pi^\pm \, \pi^\pm)$ asymmetries are expected to be tiny, which 
provides another useful test.

The asymmetry expression (\ref{JJTasym}) 
is based on a collinear factorization 
theorem (soft gluon contributions cancel, no Sudakov factors appear). 
Thus an analysis beyond tree level is
conceptually straightforward. 
The evolution of $\delta \hat{q}_I(z)$ equals that of $H_1(z)$ and is 
known to NLO Ref.\ \cite{Stratmann-Vogelsang-01}.
A NLO analysis is thus feasible (cf.\ Ref.\ \cite{Contog}).

\subsection{IFF in $e^+ e^-$ annihilation}

\noindent
For the extraction of 
the interference fragmentation functions themselves one can study a 
$\cos(\phi_{R_{1T}}^e + \phi_{R_{2T}}^e)$ asymmetry \cite{Artru-Collins} in   
$e^+ \, e^- \, \to \, (\pi^+ \, \pi^-)_{{\text{jet} \, 1}} \, (\pi^+ \,
\pi^-)_{{\text{jet} \, 2}} \, X$ which is proportional to $(\delta\hat
q_I)^{2}$ and which is again possible at BELLE. There is no 
expected asymmetric background. Combining such a result with for instance  
the single spin asymmetry in $p\, p^\uparrow \to \pi^+ \pi^- \, X$ to be 
measured at RHIC, seems --at present-- to be one of the
most realistic ways of obtaining information on the transversity function. 

Finally, a cross-check observable that is interesting to measure 
is a $\sin(\phi_{\pi}^e + \phi_{R_{T}}^e)$ asymmetry in 
$e^+ e^- \to (\pi^\pm)_{{\text{jet} \, 1}} \, (\pi^+ \,
\pi^-)_{{\text{jet} \, 2}} \, X$, which is 
proportional to a product of the Collins 
and the interference fragmentation function.  

\section*{Acknowledgments}

\noindent
I thank Matthias Grosse Perdekamp, Bob Jaffe, Jens Soeren Lange, Akio
Ogawa, Werner Vogelsang for numerous useful discussions, especially during my
employment at the RIKEN-BNL Research Center. 
At present the research of D.B. has been made possible by a 
fellowship of the Royal Netherlands Academy of Arts and Sciences.

\end{document}